\documentclass[preprint,preprintnumbers,amsmath,amssymb]{revtex4}
\usepackage{graphicx}
\usepackage{epstopdf}
\usepackage{dcolumn}
\usepackage{bm}
\usepackage{color}

\begin{document}

\title{Local electronic properties of the graphene-protected giant Rashba-split BiAg$_2$ surface}

\author{Julia Tesch,$^1$ Elena Voloshina,$^{2,}$\footnote{E-mail: elena.voloshina@hu-berlin.de} Milan Jubitz,$^1$ Yuriy Dedkov,$^{1,}$\footnote{E-mail: yuriy.dedkov@uni-konstanz.de} and Mikhail Fonin$^{1,}$\footnote{E-mail: mikhail.fonin@uni-konstanz.de}}

\affiliation{$^1$Fachbereich Physik, Universit\"at Konstanz, 78457 Konstanz, Germany}
\affiliation{$^2$Humboldt-Universit\"at zu Berlin, Institut f\"ur Chemie, 10099 Berlin, Germany}

\date{\today}

\begin{abstract}
We report the preparation of the interface between graphene and the strong Rashba-split BiAg$_2$ surface alloy and investigatigation of its structure as well as the electronic properties by means of scanning tunneling microscopy/spectroscopy and density functional theory calculations. Upon evaluation of the quasiparticle interference patterns the unpertrubated linear dispersion for the $\pi$ band of $n$-doped graphene is observed. Our results also reveal the intact nature of the giant Rashba-split surface states of the BiAg$_2$ alloy, which demonstrate only a moderate downward energy shift upon the presence of graphene. This effect is explained in the framework of density functional theory by an inward relaxation of the Bi atoms at the interface and subsequent delocalisation of the wave function of the surface states. Our findings demonstrate a realistic pathway to prepare a graphene protected giant Rashba-split BiAg$_2$ for possible spintronic applications.
\end{abstract}

\maketitle

Graphene has attracted much attention due to its unique transport, electronic and elastic properties~\cite{Geim:2007a,CastroNeto:2009,Geim:2009}. Taking into account these characteristics, many practical applications of graphene have been proposed. The most promising are graphene-based touch screens, which will potentially replace indium-tin-oxide (ITO) based screens in the future~\cite{Bae:2010,Ryu:2014fo}, batteries and supercapacitors~\cite{Wang:2009ic,Kim:2014bp,ElKady:2016fw}, and composite materials~\cite{Stankovich:2006,Huang:2012ka}. Above that a single atom thick graphene layer can effectively protect the underlying material against oxidation and/or corrosion~\cite{Sutter:2010bx,Nine:2015hka}. This property is particularly exciting when graphene is deposited or formed on the surface of a ferromagnet or a material which exhibits strong spin-orbit interaction~\cite{Dedkov:2008d,Dedkov:2008e,Dlubak:2012fq,Varykhalov:2012ec,Varykhalov:2015eb,Wang:2015aa,Farrell:2016aa}. Here the interfacial contact between graphene and the respective material might lead to the appearance of different new phenomena in graphene and at the interface, such as induced magnetism in graphene~\cite{Weser:2010,Weser:2011,Usachov:2015kr}, possible induced spin-orbit splitting of the graphene $\pi$ states~\cite{Marchenko:2012gj,Shikin:2013fr}, conservation of the spin-polarized electron emission from the underlying ferromagnetic material~\cite{Dedkov:2008d,Dlubak:2012fq}, etc.

Previously published works on the adsorption of graphene on the surfaces of heavy materials, such as Ir(111) and Au(111), demonstrate that such contacts only weakly modify the dispersion of the spin-orbit split surface states of the metal surface. Adsorption of graphene merely leads to a rigid shift of the respective surface states to smaller binding energies~\cite{Varykhalov:2012ec,Leicht:2014jy,Tesch:2016bd}, that was explained by the stronger localisation of the surface state wave function, leading to the corresponding energy shift. At the same time the intercalation of Au in the graphene/Ni(111) interface leads to the appearing of the induced spin-orbit splitting of the graphene $\pi$ states (up to $\approx100$\,meV) as a result of the hybridization of these states and valence band states of the underlying heavy metal~\cite{Marchenko:2012gj,Shikin:2013fr}. Here the energetically unfavourable model of the diluted Au atoms underneath graphene on Ni(111) was proposed~\cite{Marchenko:2012gj}. However, recent scanning tunnelling microscopy/spectroscopy (STM/STS) experiments have not shown any hints on such a splitting~\cite{Leicht:2014jh}, leaving the question of induced spin-orbit interaction in graphene still open. In order to resolve this controversy and to evaluate the role of the substrate, further experiments on graphene, which is adsorbed on materials demonstrating strong spin-orbit interaction, are required.

Here we report the fabrication of protective graphene layers on the BiAg$_2$ surface alloy, which exhibits strongly Rashba-split surface states~\cite{Ast:2007gk,ElKareh:2013gs,Schirone:2015cy}. In this system diluted Bi atoms form a $(\sqrt{3}\times\sqrt{3})R30^\circ$ structure. STM and STS used in the experiment allow to carefully separate the quasiparticle interference (QPI) signatures arising from BiAg$_2$ and from graphene that gives a possibility to to map the electronic structure of the graphene/BiAg$_2$ system around the Fermi level ($E_F$) in great detail. Our results show that the adsorption of graphene on the surface of the alloy only leads to a downward energy shift of the surface state without modifying its spin texture. We found that the $\pi$ band of the $n$-doped graphene layer has a linear dispersion in the vicinity of the Dirac point ($E_{\mathrm{D}}$) with $E_D=-400\pm30$\,meV. Contrary to the results obtained in Ref.~\citenum{Marchenko:2012gj}, our experimental and theoretical observations for graphene adsorbed on BiAg$_2$ rule out a considerable spin-orbit splitting of the graphene-derived $\pi$ states around the $K$ point ($<18$\,meV). All experiments are systematically analysed within the framework of the density functional theory (DFT) approach for the realistic experimental geometry allowing to acquire a profound understanding of the observed phenomena and discuss the possible use of this interface in future graphene-based spintronics applications. For details of the sample preparation, STM/STS experiments, and DFT calculations see Supplemental Material~\cite{suppl}.

The sequence of preparation steps during synthesis of the gr/BiAg$_2$ system (gr = graphene) is shown in Fig.\,1 (see also Fig.\,S1 of the Supplementary material for large scale STM images). In the first step, graphene nano flakes (GNFs) of different sizes are prepared by means of temperature programmed growth (TPG) procedure from C$_2$H$_4$ on Ir(111) as described elsewhere~\cite{Coraux:2009,Leicht:2014jy}. Subsequent deposition of $\approx70$\,\AA\ of Ag on GNFs/Ir(111) and annealing of this system at $450^\circ$\,C for $30$\,min leads to the formation of the GNFs/Ag(111)/Ir(111) system with graphene flakes floating on top. Earlier experimental results~\cite{Tesch:2016bd} and the present STM/STS data confirm the high quality of such a system. In the next step an almost stoichiometric BiAg$_2$ alloy is prepared underneath GNFs via adsorption of Bi atoms on GNFs/Ag/Ir(111) at $200^\circ$\,C and subsequent annealing at the same temperature for $30$\,min. This procedure leads to an effective intercalation of Bi atoms underneath the graphene layer, thus the BiAg$_2$ alloy with a $(\sqrt{3}\times\sqrt{3})R30^\circ$ structure with respect to Ag(111) is formed. The high quality of the formed GNFs/BiAg$_2$ is confirmed by the large scale STM and low-energy electron diffraction (LEED) experiments, where in the later images one can clearly resolve diffraction spots from the graphene and BiAg$_2$ subsystems as well as the respective interference picture formed by these two sub-lattices. Incomplete coverage of the graphene layer in our prepared systems (see also Fig.\,3(a)) allows to carefully trace all changes in the electronic structure of the underlying metallic layer before and after graphene adsorption as well as the corresponding modifications of the electronic properties of graphene with respect to those of a free-standing graphene layer.

In order to study the electronic band structure of the GNFs/BiAg$_2$ system, we performed combined STM/STS experiments (Figs.\,2 and 3). A large scale atomically resolved STM image and the respective $dI/dV$ map of gr/BiAg$_2$ are presented in Figs.\,2(a,b). Graphene forms a so-called moir\'e structure on the surface of the BiAg$_2$ alloy (see also Fig.\,1), which has a $(7\times 7)$ periodicity with respect to the graphene unit cell that corresponds to the $(6\times 6)$ periodicity of the Ag(111) layer where Bi atoms form a $(\sqrt{3}\times\sqrt{3})R30^\circ$ structure [Fig.\,4(a)]. A Fast-Fourier transform (FFT) of the $dI/dV$ map yields the image presented in Fig.\,2(c), where several characteristic features can be identified. The first one, which is marked by the dashed line hexagon, is assigned to the reciprocal lattice of a graphene layer [reciprocal vector $\vec{b}_{gr}$ in Fig.\,2(c)]. Each central spot here is surrounded by six additional ones, which are due to the moir\'e lattice formed at the gr/BiAg$_2$ interface. Bi atoms, which form a $(\sqrt{3}\times\sqrt{3})R30^\circ$ overstructure of the BiAg$_2$ alloy, are responsible for the respective spots in the FFT map [reciprocal vector $\vec{b}_{Bi}$ in Fig.\,2(c)]. Analogously to the previous discussion, the additional six spots are resolved as well.

The most interesting features in the FFT can be identified around â$\vec{q} = 0$ as well as around the $(\sqrt{3}\times\sqrt{3})R30^\circ$ points with respect to the graphene atomic lattice. Corresponding areas are marked by the solid line rectangles, and their zooms are shown in Figs.\,2 (d) and (e), respectively. The discussed features in the FFT maps obtained from the $dI/dV$ images acquired at different bias voltages ($U_T$) can be assigned to QPI patterns formed after scattering of the electron waves at the surface defects (steps, dislocations, adatoms, etc.). Analysis of such maps allows to identify particular scattering vectors ($q_E$) between different electronic states in the Brillouin tone (BZ) at the fixed energy, $E=eU_T$, and to plot the energy dispersion of the carriers $E(k)$ around $E_F$.

The feature around â$\vec{q} = 0$ shown in Fig.\,2(d) is formed by the scattering vectors connecting different Rashba-split surface states of BiAg$_2$ and is very similar to the one observed earlier~\cite{ElKareh:2013gs,Schirone:2015cy} (see Fig.\,S2 of the Supplementary material for the series of FFT maps around $\vec{q} = 0$ obtained for gr/BiAg$_2$ at different $U_T$). Analysis performed in Refs.~\citenum{ElKareh:2013gs,Schirone:2015cy} allows to identify them as $D_1$, $D_2$, and $D_3$ (notation is according to Ref.~\citenum{ElKareh:2013gs}). They are marked by the respective symbols in Fig.\,S3 of the Supplementary material, where the DFT calculated band structure of the BiAg$_2$/Ag(111) slab is shown. The zoom of the FFT presented in Fig.\,2(e) reveals a ``ring-like'' structure. It is assigned to the intervalley scattering between graphene-related valence band states around the $\mathrm{K}$ and $\mathrm{K'}$ points. The radius of these rings is $2k$, where $k$ is the wave vector of the Dirac particles at an energy $E$ relative to $E_F$ and it is measured with respect to the $\mathrm{K}$ point of the graphene BZ. The deviation of the shape of these ``ring-like'' features from the circle for the large positive bias voltages is due to the trigonal warping at the energies far away from $E_D$ as was shown in theoretical calculations~\cite{Simon:2011dv}.

Figure\,3 shows the extracted dispersions of the wave vector $k$ gathered from a series of $dI/dV$ maps measured for BiAg$_2$ and GNF/BiAg$_2$ as discussed above (see STM image of the border between two regions in panel (a)): (b) and (c) are for the Rashba-split surface states of BiAg$_2$ around the $\Gamma$ point without and with a graphene layer on top, respectively, and (d) is for the graphene $\pi$ states around the $\mathrm{K}$ point of the graphene-derived BZ. Analysis of the graphene dispersion relation shows that graphene on BiAg$_2$ is $n$-doped with a position of the Dirac point of $E_D=-400\pm 30$\,meV and Fermi velocity of $(1.17\pm 0.06)\times10^6$\,m/s, which is in good agreement with a value for nearly free-standing graphene on a metallic substrate~\cite{Trevisanutto:2008aa,Hwang:2012aaa}. 

A peculiar behaviour is encountered upon comparison of Fig.\,3 (b) and (c), where the dispersions of the scattering vector for the surface state of BiAg$_2$ and GNF/BiAg$_2$ are presented, respectively. These data were obtained on the same sample under similar experimental conditions excluding any experimental or/and tip artefacts influencing the final result. One can clearly see that after adsorption of graphene on BiAg$_2$ all Rashba-split surface states are shifted downwards in energy by about $100-150$\,meV. A similar shift is detected for the single $dI/dV$ spectra measured as a function of $U_T$ for the neighbouring areas of BiAg$_2$ and gr/BiAg$_2$ (see Fig.\,S4 of the Supplementary material). This effect is contrary to the previously observed results for graphene adsorption on Au(111)~\cite{Leicht:2014jy,Tesch:2016bd}, Ag(111)~\cite{Jolie:2015ba,Tesch:2016bd,Garnica:2016hu}, Cu(111)~\cite{GonzalezHerrero:2016eh}, and Ir(111)~\cite{Subramaniam:2012fp,Varykhalov:2012ec}, where an upward energy shift for the metallic surface states was reported and explained by the stronger localization of the wave function of the metallic surface state producing an increase of Pauli repulsion at the interface. This effect leads to an increase of the energy of the electrons and consequently to the upward energy shift of the surface state. Moreover, recent ARPES experiments also reveal that adsorption of the atoms or molecules with closed shells like rare gas Xe, $\mathrm{C}_{60}$, FeOEP, or PTCDA on the surface of the BiAg$_2$ alloy does not lead to the energy shift of these Rashba-split states and $k$-splitting remains intact~\cite{Moreschini:2008fx,Cottin:2014iy}.

In order to fully understand all observed effects we performed modelling of the electronic properties of the gr/BiAg$_2$ system in the framework of the DFT approach. For the modelling of the surface of the BiAg$_2$ alloy, 1/3 of Ag atoms in the top layer of the 7-layers Ag(111) slab were replaced with Bi atoms forming the characteristic $(\sqrt{3}\times\sqrt{3})R30^\circ$ superstructure. A graphene layer with a $(7\times7)$ periodicity was adsorbed on one side of this slab, which respectively has a $(6\times6)$ periodicity with respect to the Ag(111) in-plane lattice. The resulting structure is shown in Fig.\,4(a,b), where the formed moir\'e with a periodicity of $17.523$\,\AA\ is clearly visible. The following systems were analysed: (\textbf{A}) clean BiAg$_2$ surface after relaxation of coordinates of the top Bi and Ag layers, (\textbf{B}) gr/BiAg$_2$-fixed where coordinates of carbon atoms were fully relaxed and coordinates of the top layer of Bi and Ag atoms were fixed as obtained for the system (\textbf{A}), and (\textbf{C}) gr/BiAg$_2$-relaxed, where coordinates of carbon atoms and the top layer of Bi and Ag atoms were fully relaxed (see Table\,T1 of the Supplementary material for the resulting interlayer distances for all considered structures). In our analysis of the electronic properties of the studied systems, the resulting band structures were unfolded on the $(1\times1)$ Brillouin zone of the respective sublattice, Ag(111) or graphene.

Figure\,5 shows the unfolded band structures around the $\Gamma_{10}^\mathrm{BiAg_2}$-point of BZ corresponding to the BiAg$_2$(111) unit cell for (a) BiAg$_2$ (system \textbf{A}), (b) gr/BiAg$_2$-fixed (system \textbf{B}), and (c) gr/BiAg$_2$-relaxed (system \textbf{C}). The choice of the $\Gamma_{10}^\mathrm{BiAg_2}$-point (which coincides with the $\mathrm{K}_{00}^\mathrm{Ag}$-point of the BZ corresponding to the Ag(111) unit cell) is caused by the appearance of the artificial Ag(111) surface state at $E-E_F=-100$\,meV originating from the back side of the slab used in the DFT computation (see Figs.\,S5 -- S9 of the Supplementary material for the additional figures). In panel (d) the band structure of graphene around the $\mathrm{K}_{00}^\mathrm{gr}$-point of the graphene-derived BZ for the gr/BiAg$_2$-relaxed (system \textbf{C}) is shown.

The performed DFT calculations for the above described systems confirm our experimental observations for the downward energy shift of the BiAg$_2$ surface state after graphene adsorption. In Fig.\,5(a) the dispersions of the Rashba-split surface states for the clean BiAg$_2$ surface are clearly resolved. The energy positions of the crossing points of the Rashba-split surface states at the $\Gamma_{10}^\mathrm{BiAg_2}$-point are $-280$\,meV, $575$\,meV, and $1395$\,meV, respectively. Adsorption of graphene on the surface of the alloy, without relaxation of the atomic positions for Bi and Ag top layers, leads to the upward energy shift of all states and the resulting energy position of the crossing points are $-225$\,meV, $585$\,meV, and $1395$\,meV, respectively (Fig.\,5(b)). This effect can be explained by the stronger localisation of the surface states wave functions that leads to an increase of Pauli repulsion for these states and thus to the energy shift to smaller binding energies. 

Relaxation of all atomic positions at the interface between graphene and BiAg$_2$ alloy, which models the real experimental situation, leads to the opposite effect compared to the previously described case (Fig.\,5(c)). Here we observed the relatively large downward energy shift of the Rashba-split surface states of BiAg$_2$. The energy positions of the crossing points at the $\Gamma_{10}^\mathrm{BiAg_2}$-point are $-355$\,meV, $515$\,meV, and $1340$\,meV, respectively. Our calculations show that adsorption of graphene on the surface of the BiAg$_2$ alloy leads to the inward Bi-atom relaxation with a reduction of the mean distance between planes formed by Bi and Ag atoms by $0.12$\,\AA. At the same time the mean distance between graphene and the top Bi layer is slightly increased from $3.219$\,\AA\ for BiAg$_2$-fixed (system \textbf{B}) to $3.277$\,\AA\ for BiAg$_2$-relaxed (system \textbf{C}). Both effects lead to the stronger delocalisation of the wave function of the surface states compared to the clean surface of the BiAg$_2$ alloy. Thus the reduction of Pauli repulsion in turn leads to the increase of the binding energies of all surface states upon graphene adsorption. The effect of delocalisation can be visualised via 2D presentation (side and top view) of the calculated electron localization function (ELF)~\cite{Becke:1990kd,Savin:1997vh,DeSantis:2000fs,Qi:2005jm}. 
Analysis of ELF presented in Fig.\,S10 of the Supplementary material confirms the effect of stronger delocalization of the surface state for the gr/BiAg$_2$-relaxed system (system \textbf{C}), compared to the previous case of the non-fully relaxed system. 

Our DFT results show that graphene is $n$-doped in both cases of its adsorption on BiAg$_2$ (fixed or relaxed structure) with a position of the Dirac point of $E_D-E_F=-545$\,meV (system \textbf{B}) and  $E_D-E_F=-590$\,meV (system \textbf{C}) (Fig.\,5(d)). This value is in reasonably good agreement with the doping level of graphene of $-400\pm30$\,meV obtained in the STM/STS experiments. Our calculations demonstrate the absence of the energy gap at $E_D$ for gr/BiAg$_2$-fixed and opening of the gap for the graphene $\pi$ states of $\approx95$\,meV at $E_D$ for the gr/BiAg$_2$-relaxed system. Despite the presence of the substrate underneath graphene, which demonstrates strong spin-orbit interaction, no spin-splitting for the graphene-derived $\pi$ states is found in experiment as well as in the DFT calculations. This can be explained by the relatively large gr-Bi distance of $\approx3.3$\,\AA\ in the studied systems. These results are also confirmed by our STM/STS experiments where no evidence of the splitting of the graphene states leading to modifications of the relevant scattering features was observed. However, we believe that further studies by means of ARPES (spin-ARPES) might shed more light on the possible existence of the energy gap in the graphene band structure as well as on the spin-texture of the valence band states in the vicinity of $E_F$ for the gr/BiAg$_2$ system.

In summary, we demonstrate the successful preparation of graphene on the strong Rashba-split BiAg$_2$ surface alloy via subsequent intercalation of Ag and Bi in graphene/Ir(111). Our method grants large flexibility in the fabrication of big and small graphene flakes in this system. Presented systematic STM/STS experiments show that graphene on BiAg$_2$ is $n$-doped with a position of the Dirac point of $-400\pm30$\,meV below $E_F$ and that the $\pi$ states have a linear dispersion confirming the massless nature of carriers in the vicinity of the Fermi level. In contrast to the previously studied cases of a graphene adsorption on noble metal surfaces, here we observed the downward shift of the giant Rashba-split surface states of the BiAg$_2$ surface alloy for the graphene covered surface compared to the clean one, rendering the spin-texture of the graphene-protected BiAg$_2$ surface alloy unaffected by graphene adsorption. Our experimental findings were analysed in the framework of the DFT approach where an inward relaxation of the Bi atoms in BiAg$_2$ upon graphene adsorption was found, consequently leading to the delocalization of the surface state wave function. Despite the presence of the strong Rashba-split BiAg$_2$ surface alloy in contact with graphene, no sizeable spin-splitting (above $18$\,meV) was detected for the graphene $\pi$ states as confirmed by our STM/STS and DFT results. The studied system, where a lattice of diluted Bi atoms is placed underneath graphene on Ag(111), is a perfect model example demonstrating the strong Rshba-splitting of the surface states and the absence of the induced spin-orbit splitting in graphene. The presented results solve the controversy on the possible observation of the spin-splitting phenomena in graphene physisorbed on the metallic surfaces with large Rashba-splitting.

We thank the German Research Foundation (DFG) for financial support within the Priority Programme 1459 ``Graphene'' and the North-German Supercomputing Alliance (HLRN) for providing computer time.


\clearpage
\begin{figure}
\center
\includegraphics[width=\textwidth]{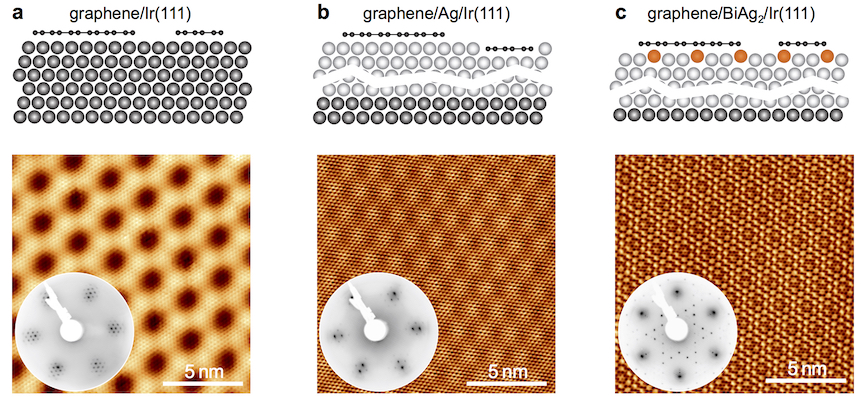}\\
\vspace{1cm}
\caption{(a) GNFs/Ir(111), (b) GNFs/Ag(111)/ Ir(111), and (c) GNFs/BiAg$_2$/Ir(111). Bottom row shows corresponding STM and LEED images acquired after every preparation step. Imaging parameters: (a) $U_T=50$\,mV, $I_T=500$\,pA; (b) $U_T=-150$\,mV, $I_T=800$\,pA; (c) $U_T=50$\,mV, $I_T=700$\,pA. LEED images were obtained at $75$\,eV.}
\end{figure}

\clearpage
\begin{figure}
\center
\includegraphics[width=\textwidth]{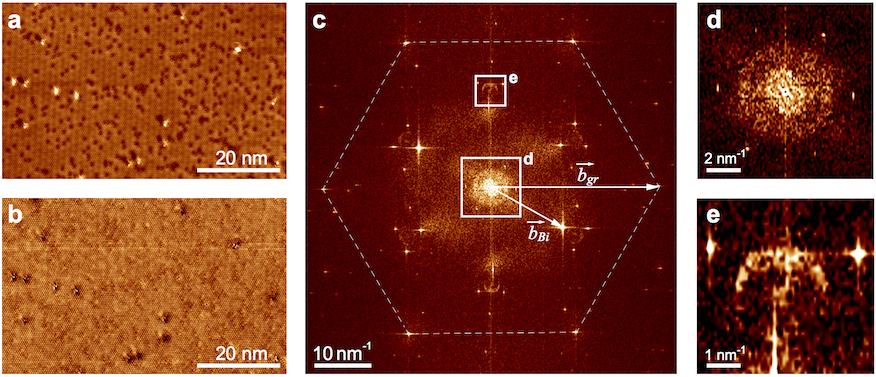}\\
\vspace{1cm}
\caption{(a) Topographic, $z(x,y)$, and (b) differential conductance, $dI/dV(x,y)$, maps acquired on GNF/BiAg$_2$. Imaging parameters: $U_T=+125$\,mV, $I_T=900$\,pA, $f_{mod}=687$\,Hz, $U_{mod}=10$\,mV. (c) FFT image obtained from (b). Dashed hexagon marks the reciprocal lattice of graphene with vector $\vec{b}_{gr}$. Vector $\vec{b}_{Bi}$ marks reciprocal lattice of BiAg$_2$. Rectangles are used for areas around the $\Gamma$ point and the $\mathrm{K}$ point of the graphene BZ, zooms of which are shown in (d) and (e), respectively.}
\end{figure}

\clearpage
\begin{figure}
\center
\includegraphics[width=\textwidth]{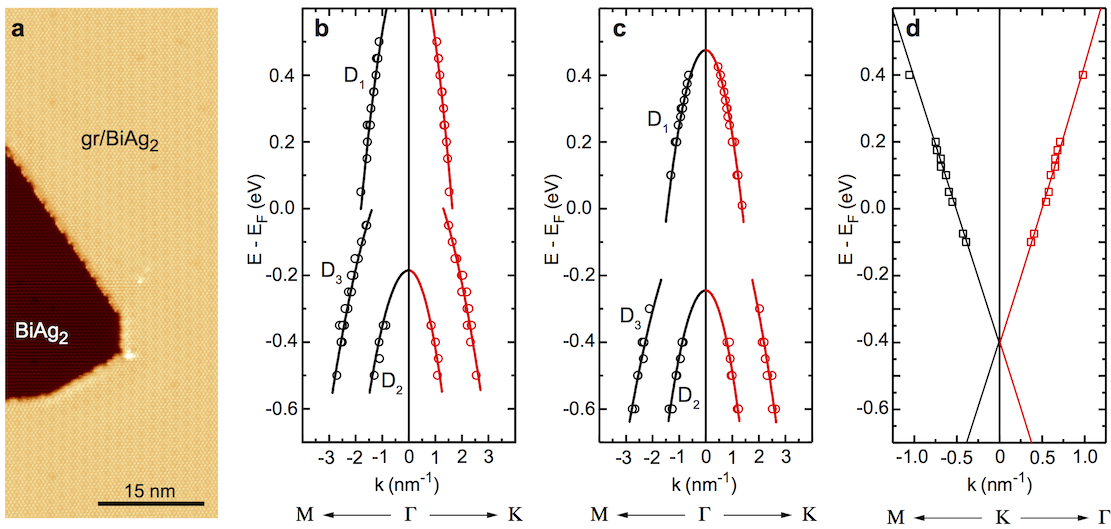}\\
\vspace{1cm}
\caption{(a) STM topography for the simultaneous imaging of areas of BiAg$_2$ and gr/BiAg$_2$. Imaging parameters: $75\times30\mathrm{nm}^2$, $U_T=20$\,mV, $I_T=2$\,nA. Scattering vector dispersions $k(E)$ obtained at different $U_T$ corresponding to $D_1$, $D_2$, $D_3$ for (b) clean BiAg$_2$ and (c) gr/BiAg$_2$. (d) scattering vector dispersion corresponding to the intervalley scattering in graphene for gr/BiAg$_2$.}
\end{figure}

\clearpage
\begin{figure}
\center
\includegraphics[width=0.75\textwidth]{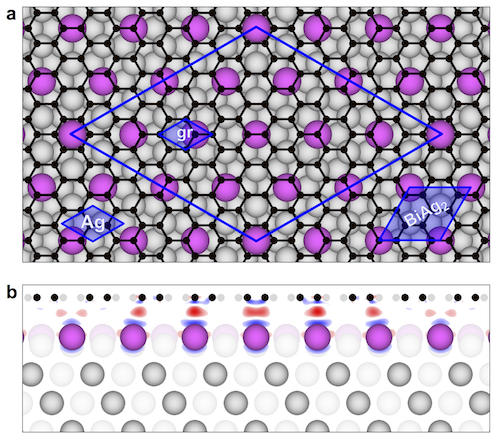}\\
\vspace{1cm}
\caption{Top (a) and side (b) views of the gr/BiAg$_2$ interface. In (b) the charge difference, $\Delta\rho_{gr/sub}(r)=\rho_{gr/sub}(r)-(\rho_{gr}(r)+\rho_{sub}(r))$ is color coded -- from red ($+2\,e/\mathrm{nm}^3$) to blue ($-2\,e/\mathrm{nm}^3$).}
\end{figure}

\clearpage
\begin{figure}
\center
\includegraphics[width=\textwidth]{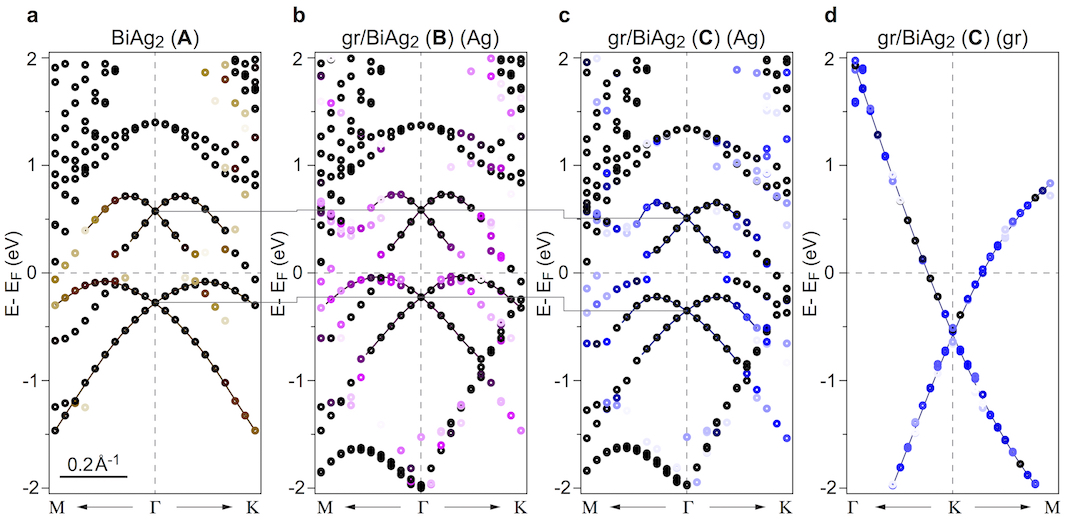}\\
\vspace{1cm}
\caption{Calculated energy band dispersions for (a) clean BiAg$_2$ (system \textbf{A}), (b) gr/BiAg$_2$-fixed (system \textbf{B}), (c-d) gr/BiAg$_2$-relaxed (system \textbf{C}).}
\end{figure}

\clearpage

\noindent
Supplementary material for manuscript:\\
\textbf{Local electronic properties of the graphene-protected giant Rashba-split BiAg$_2$ surface}\\
\newline
Julia Tesch,$^1$ Elena Voloshina,$^2$ Milan Jubitz,$^1$ Yuriy Dedkov,$^{1}$ and Mikhail Fonin$^1$\\
\newline
$^1$Fachbereich Physik, Universität Konstanz, 78457 Konstanz, Germany\\
$^2$Humboldt-Universit\"at zu Berlin, Institut f\"ur Chemie, 10099 Berlin, Germany\\
\\
\noindent
\textbf{Sample preparation.} The Ir(111) crystal (MaTecK GmbH and SPL) used as a substrate was cleaned by several cycles of Ar$^+$ sputtering ($2$\,keV), oxygen annealing ($900-1150^\circ$\,C, $5\times10^{-7}$\,mbar O$_2$ pressure) and flash annealing ($5$\,s up to $1800^\circ$\,C). The cleanliness of the Ir(111) crystal was checked by LEED and STM. Graphene was prepared by $1-2$ subsequent temperature programmed growth (TPG) cycles. For each TPG cycle the Ir(111) surface was exposed to an ethylene dose of $6.8$ Langmuir at room temperature and subsequently annealed for $25$\,s with the temperature held between $1100^\circ$\,C and $1330^\circ$\,C. Silver ($\approx70$\,\AA) was evaporated from an effusion cell. Subsequently, the sample was annealed at $450^\circ$\,C for $30$\,min leading to the formation of the GNFs/Ag(111)/Ir(111) system. In the next step the close to stoichiometric BiAg$_2$ alloy underneath GNFs is prepared via adsorption of Bi atoms on gr/Ag/Ir(111) at $200^\circ$\,C and subsequent annealing at the same temperature for $30$\,min. This procedure leads to the effective penetration of Bi atoms underneath the graphene layer and the BiAg$_2$ alloy with a $(\sqrt{3}\times\sqrt{3})R30^\circ$ structure with respect to Ag(111) is formed. The high quality of the formed GNFs/BiAg$_2$ is confirmed by the large scale STM and LEED measurements.

\noindent
\textbf{STM/STS experiments.} STM and STS measurements were performed at $5.5$\,K in an Omicron Cryogenic STM under ultra-high vacuum (UHV) conditions ($<5\times10^{-11}$\,mbar). Polycrystalline UHV-flash-annealed tungsten and PtIr tips were used for all STM/STS measurements. The sign of the bias voltage corresponds to the potential applied to the sample. Differential conductance ($dI/dV$) maps were recorded by means of standard lock-in technique, using the modulation voltages and frequencies given in the figure captions.

\noindent
\textbf{DFT calculations.} DFT calculations based on plane-wave basis sets of $500$\,eV cutoff energy were performed with the Vienna \textit{ab initio} simulation package (VASP)~\cite{Kresse:1994,Kresse:1996a,Kresse:1999}. The Perdew-Burke-Ernzerhof (PBE) exchange-correlation functional~\cite{Perdew:1996} was employed. The electron-ion interaction was described within the projector augmented wave (PAW) method~\cite{Blochl:1994} with Ag ($4d$, $5s$), Bi ($6s$, $6p$), and C ($2s$, $2p$) states treated as valence states. The Brillouin-zone integration was performed on $\Gamma$-centred symmetry reduced Monkhorst-Pack meshes using a Methfessel-Paxton smearing method of first order with $\sigma = 0.2$\,eV, except for the calculations of total energies. For those calculations, the tetrahedron method with Bl\"ochl corrections~\cite{Blochl:1994vg} was used.  A $3\times 3\times 1$  $k$-mesh was used in the case of ionic relaxations and $6\times 6\times 1$ for single point calculations, respectively. Dispersion interactions were considered adding a $1/r^6$ atom-atom term as parameterised by Grimme (``D2'' parameterisation)~\cite{Grimme:2006}. The spin-orbit correction, necessary for the proper description of the properties of the BiAg$_2$ surface alloy, is taken into account via non-collinear magnetism as implemented in VASP.

The supercell used in this work has a ($6\times 6$) lateral periodicity with respect to Ag(111). It is constructed from a slab of 6 layers of Ag, one interface layer of BiAg$_2$ and one layer of graphene adsorbed on one side and a vacuum region of approximately $23$\,\AA\ (Fig.\,4(a,b)). The positions ($x$, $y$, $z$-coordinates) of the ions of the top 2 layers (i.\,e. graphene and BiAg$_2$) as well as $z$-coordinates of the third layer (i.\,e. Ag) were fully relaxed until forces became smaller than $0.02$\,eV\,\AA$^{-1}$. The lattice constant in the lateral plane was set according to the optimised value of the \textit{fcc} Ag, $a_{\textrm{hex}} = 4.13/\sqrt{2}=2.92$\,\AA. The lattice mismatch between graphene and substrate in this case is approximately $1.5$\,\%.  

The band structures calculated for the studied systems were unfolded to the graphene ($1\times 1$) and Ag ($1\times 1$) primitive unit cells according to the procedure described in Refs.~\citenum{Medeiros:2014ka,Medeiros:2015ks} with the code BandUP.


\newpage
{\def\arraystretch{1.2}\tabcolsep=24pt
\begin{tabular}{ l | c | c | c }
\hline
             &clean BiAg$_2$&\multicolumn{2}{c}{gr/BiAg$_2$} \\
\cline{3-4}
Distance     &       & Fixed & Relaxed \\
\hline
gr -- Bi      &       &$3.219$&$3.277$\\
Bi(S) -- Ag(S)     &$0.740$&$0.740$&$0.623$\\
Ag(S) -- Ag(S-1)&$2.387$&$2.387$&$2.434$\\
             &       &       &       \\
gr-corrugation      &       &$0.019$&$0.013$\\
Bi-corrugation      &$0.000$&$0.000$&$0.020$\\ 
\hline
\end{tabular}
}\\
\newline
\noindent\textbf{Table\,T1.} Interlayer mean distances between planes of graphene, Bi atoms, and Ag atoms (in \AA) for clean BiAg$_2$ (system \textbf{A}), gr/BiAg$_2$-fixed (system \textbf{B}), and gr/BiAg$_2$-relaxed (system \textbf{C}).

\newpage
\begin{figure}[h]
  \includegraphics[width=\textwidth]{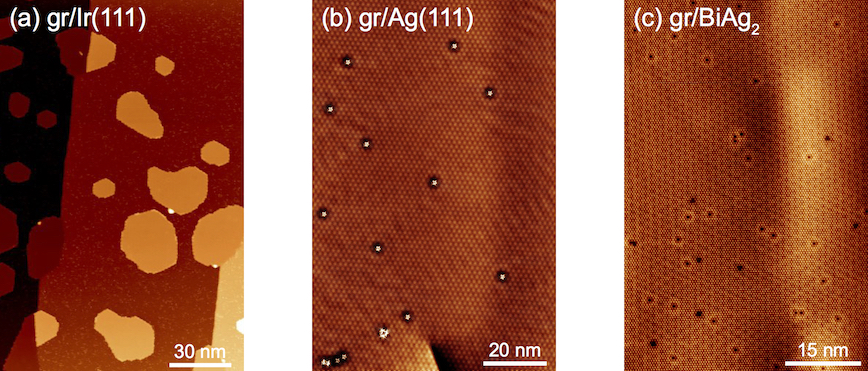}\\
\label{S1_SUPPL_large_scale_STM}
\end{figure}
\noindent\textbf{Fig.\,S1.} Large scale STM images of (a) GNFs/Ir(111), $U_T=+300$\,mV, $I_T=1.6$\,nA, (b) gr/Ag(111)/Ir(111), $U_T=+10$\,mV, $I_T=0.8$\,nA, (c) gr/BiAg$_2$/Ir(111), $U_T=+100$\,mV, $I_T=0.7$\,nA.

\newpage
\begin{figure}[h]
  \includegraphics[width=\textwidth]{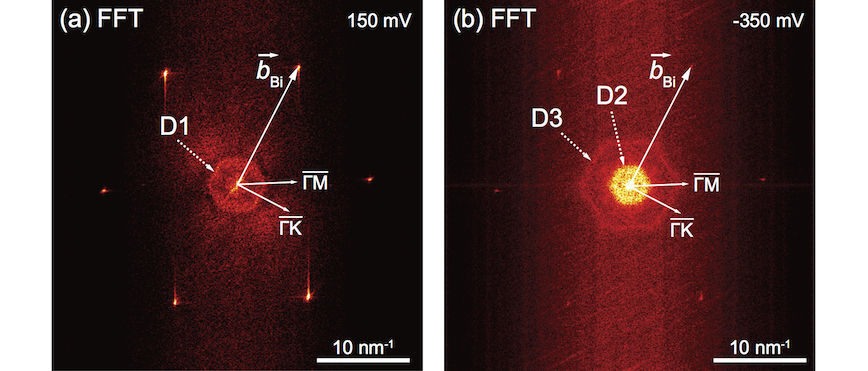}\\
\label{S2_SUPPL_FFT_maps}
\end{figure}
\noindent\textbf{Fig.\,S2.} Two representative FFT images obtained on the basis of the $dI/dV$ maps acquired on BiAg$_2$ at (a) $U_T=+150$\,mV and (b) $U_T=-350$\,mV. The respective signals assigned to $D_1$, $D_2$, and $D_3$ are marked in the images. $\vec{b}_{Bi}$ is the vector of the reciprocal lattice of BiAg$_2$ and  $\Gamma\mathrm{K}$ and $\Gamma\mathrm{M}$ are two directions in the corresponding BZ. 

\newpage
\begin{figure}[h]
  \includegraphics[width=0.75\textwidth]{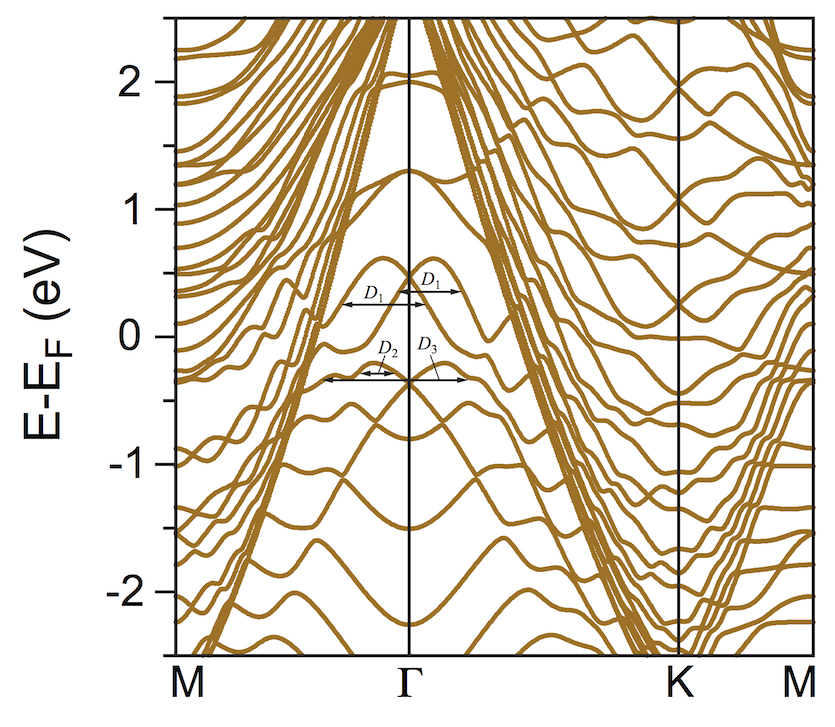}\\
\label{S3_SUPPL_band_str_BiAg2_Ag111}
\end{figure}
\noindent\textbf{Fig.\,S3.} Band structure of the BiAg$_2$ surface alloy along several high symmetry directions of the BZ. $D_1$, $D_2$, and $D_3$ vectors connecting the branches in the electronic structure of equal spin are marked in the figure.

\newpage
\begin{figure}[h]
  \includegraphics[width=\textwidth]{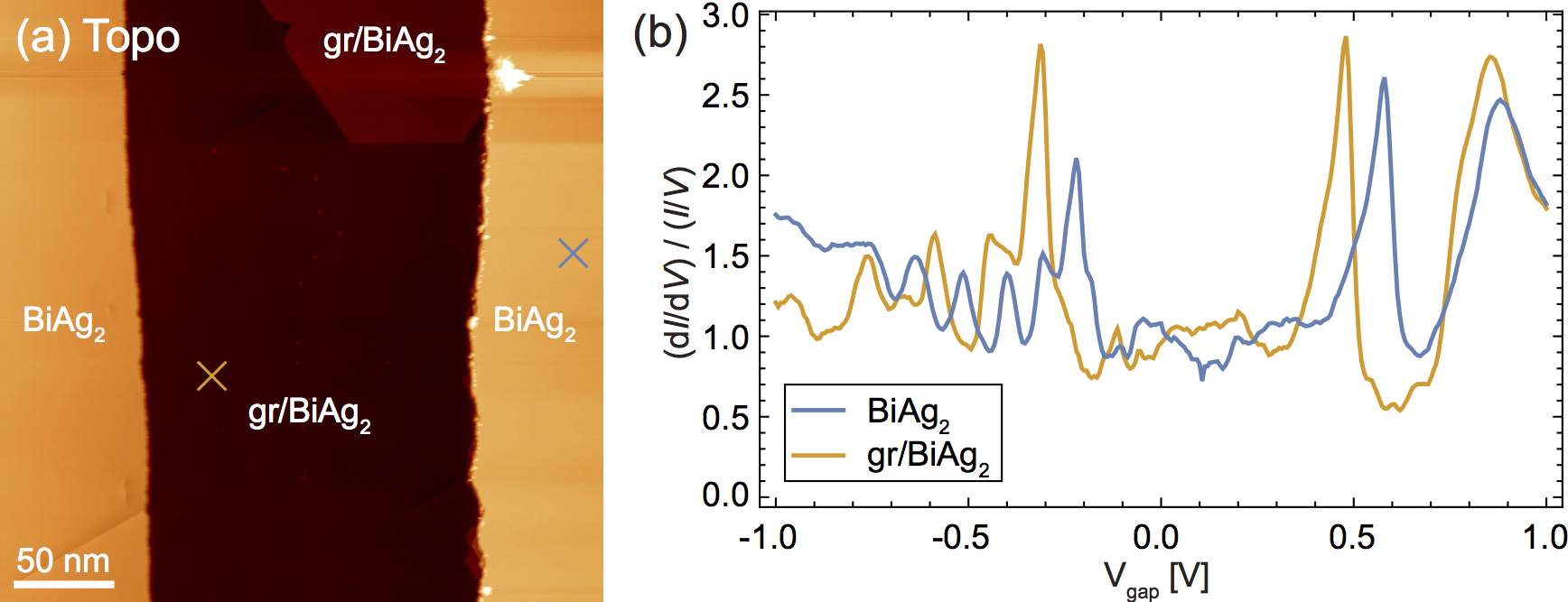}\\
\label{S4_SUPPL_dI_dV}
\end{figure}
\noindent\textbf{Fig.\,S4.} (a) STM topography of two neighbouring regions of BiAg$_2$ and gr/BiAg$_2$ ($U_T=+50$\,mV, $I_T=700$\,pA) and (b) representative $dI/dV$ spectra ($f_{mod}=678.2$\,Hz, $U_{mod}=6$\,mV) measured in the respective areas (exact places are marked by crosses in (a)).

\newpage
\begin{figure}[h]
  \includegraphics[width=0.75\textwidth]{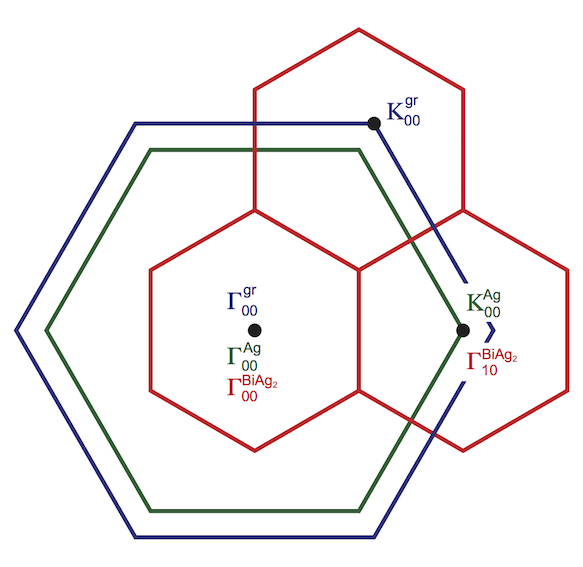}\\
\label{S5_SUPPL_BZs_scheme}
\end{figure}
\noindent\textbf{Fig.\,S5.} Schematic representation of the BZs corresponding to $(1\times1)$ unit cells and the respective high symmetry points: (red) BZ of BiAg$_2$, (green) BZ of Ag(111), (blue) BZ of graphene.

\newpage
\begin{figure}[h]
  \includegraphics[width=\textwidth]{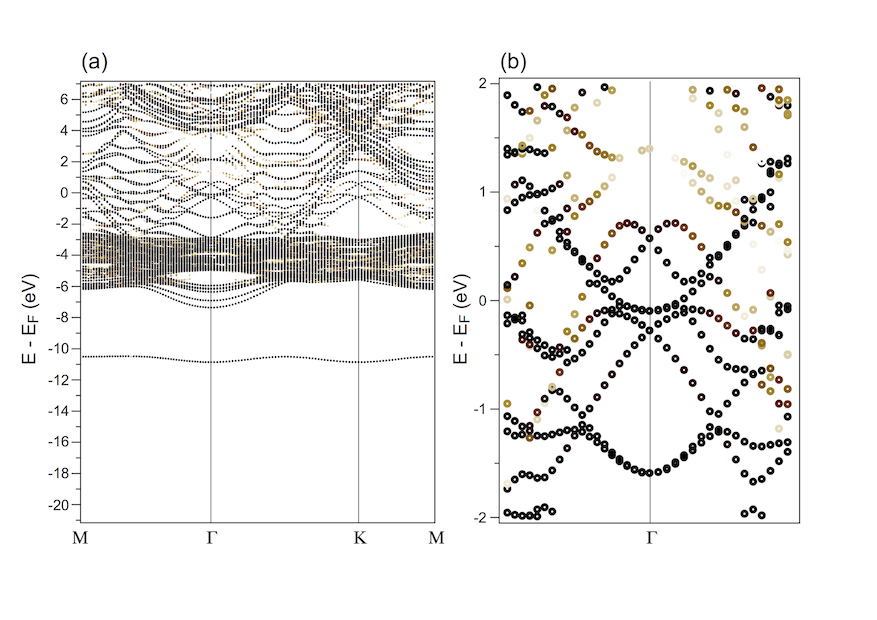}\\
\label{S6_SUPPL_band_str_BiAg2_clean}
\end{figure}
\noindent\textbf{Fig.\,S6.} (a) Band structure of the clean $(6\times6)$ BiAg$_2$ (system \textbf{A}) unfolded on the BZ of Ag(111). High symmetry points correspond to BZ of Ag(111). (b) Zoom around $\Gamma$ point.

\newpage
\begin{figure}[h]
  \includegraphics[width=\textwidth]{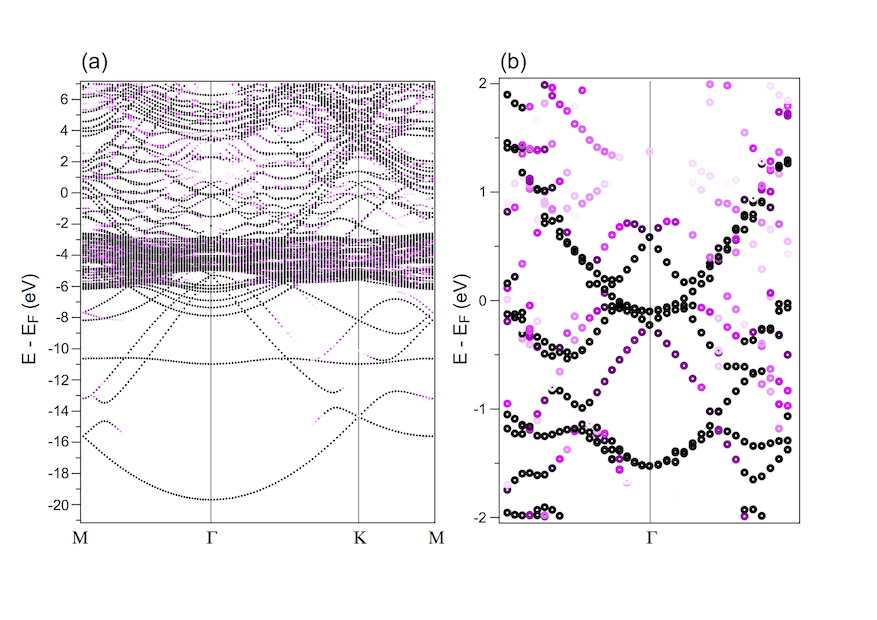}\\
\label{S7_SUPPL_band_str_grBiAg2_fixed}
\end{figure}
\noindent\textbf{Fig.\,S7.} (a) Band structure of $(7\times7)$gr/$(6\times6)$ BiAg$_2$-fixed (system \textbf{B}) unfolded on the BZ of Ag(111). High symmetry points correspond to BZ of Ag(111). (b) Zoom around $\Gamma$ point.

\newpage
\begin{figure}[h]
  \includegraphics[width=\textwidth]{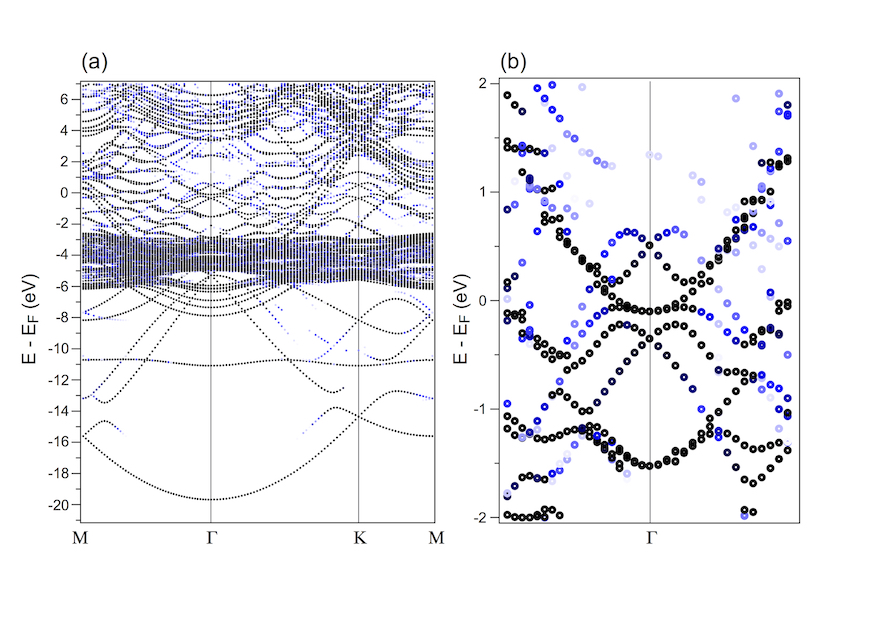}\\
\label{S8_SUPPL_band_str_grBiAg2_relaxed}
\end{figure}
\noindent\textbf{Fig.\,S8.} (a) Band structure of $(7\times7)$gr/$(6\times6)$ BiAg$_2$-relaxed (system \textbf{C}) unfolded on the BZ of Ag(111). High symmetry points correspond to BZ of Ag(111). (b) Zoom around $\Gamma$ point.

\newpage
\begin{figure}[h]
  \includegraphics[width=0.55\textwidth]{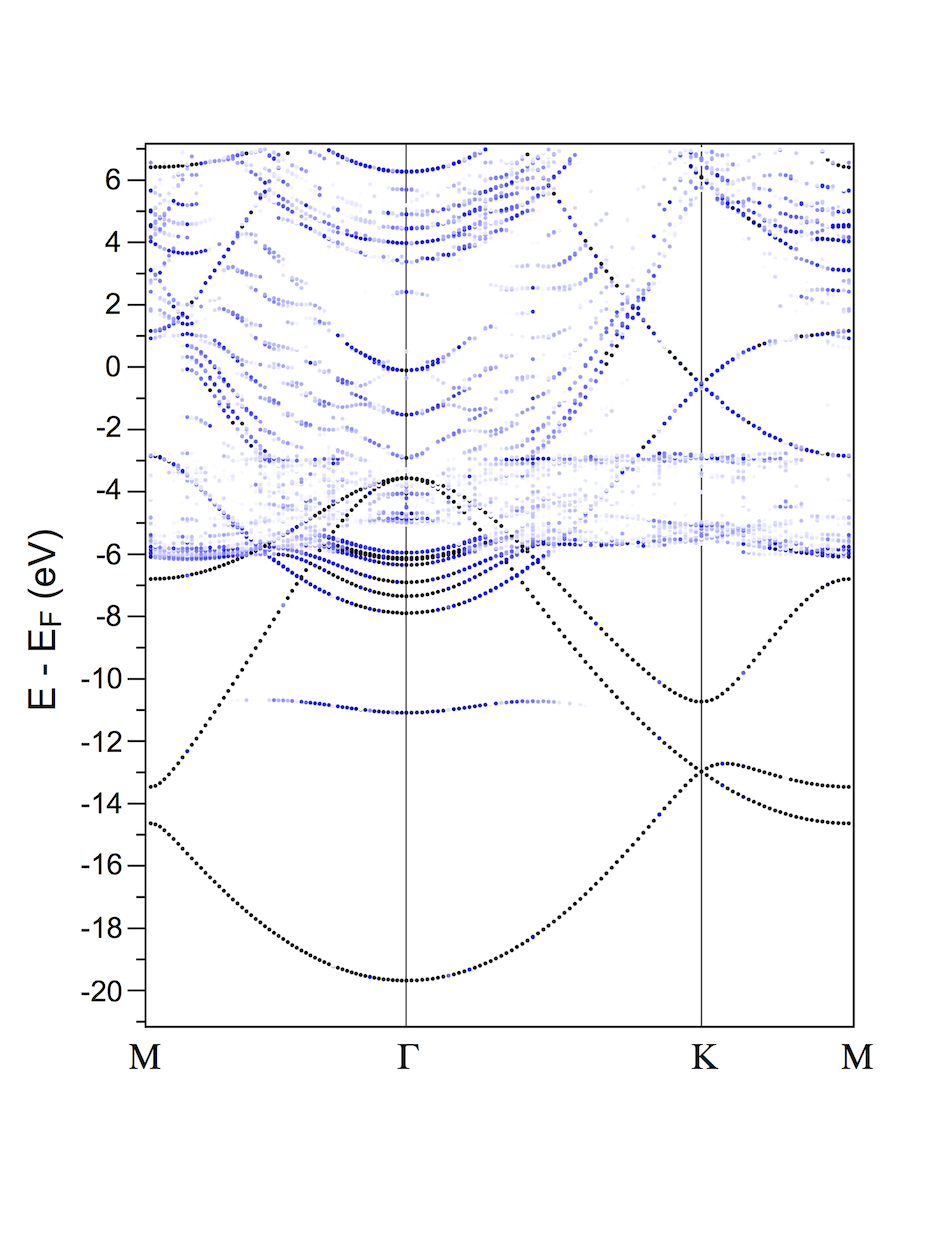}\\
\label{S9_SUPPL_band_str_graphene_grBiAg2_relaxed}
\end{figure}
\noindent\textbf{Fig.\,S9.} Band structure of $(7\times7)$gr/$(6\times6)$ BiAg$_2$-relaxed (system \textbf{C}) unfolded on the BZ of graphene. High symmetry points correspond to BZ of graphene.

\newpage
\begin{figure}[h]
  \includegraphics[width=\textwidth]{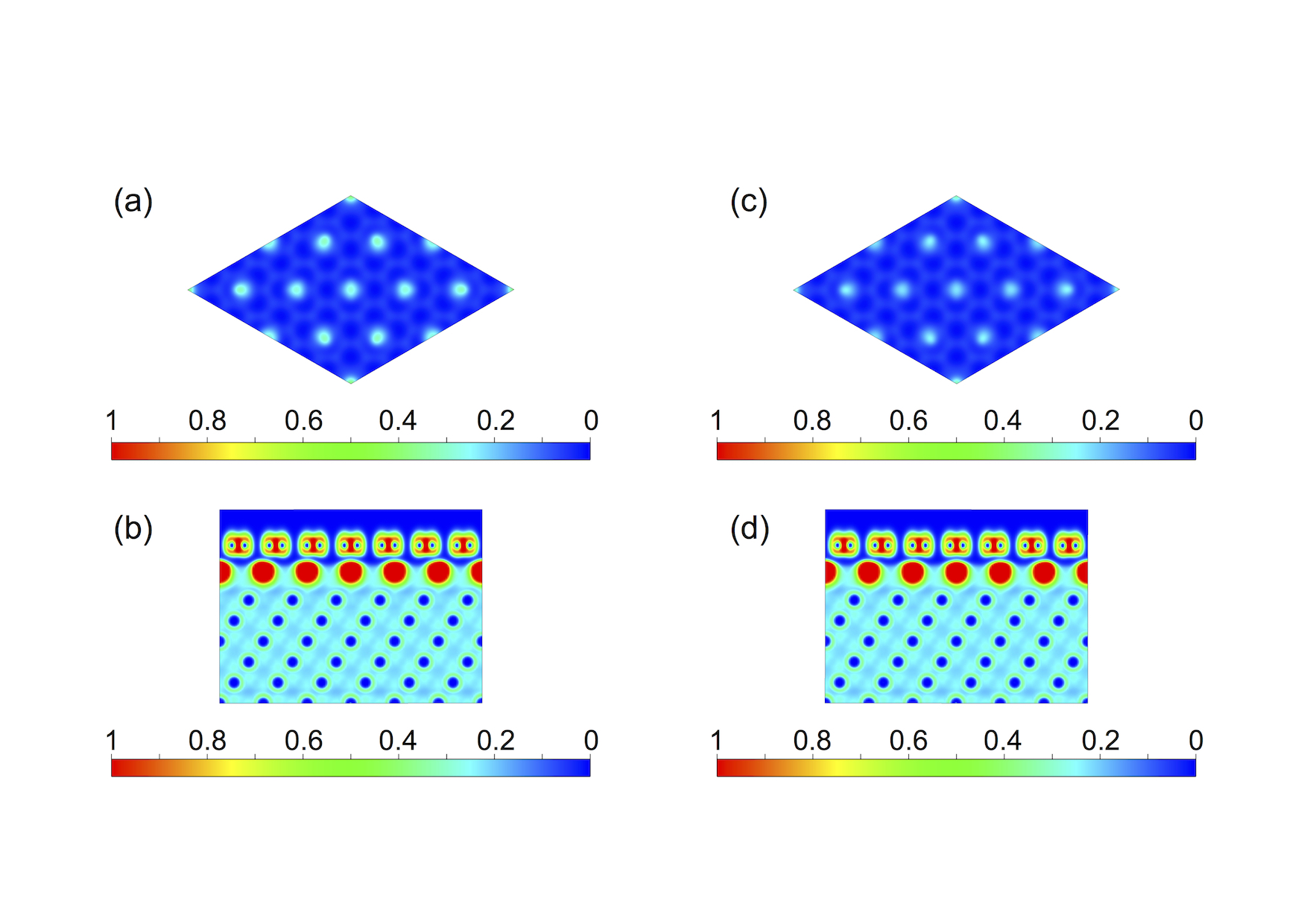}\\
\label{S10_SUPPL_ELF_grBiAg2_fix_relax}
\end{figure}
\noindent\textbf{Fig.\,S10.} Top and side views of ELF function for (a,b) gr/BiAg$_2$-fixed (system \textbf{B}) and (c,d) gr/BiAg$_2$-relaxed (system \textbf{C}).

\newpage
\begin{figure}[h]
  \includegraphics[width=\textwidth]{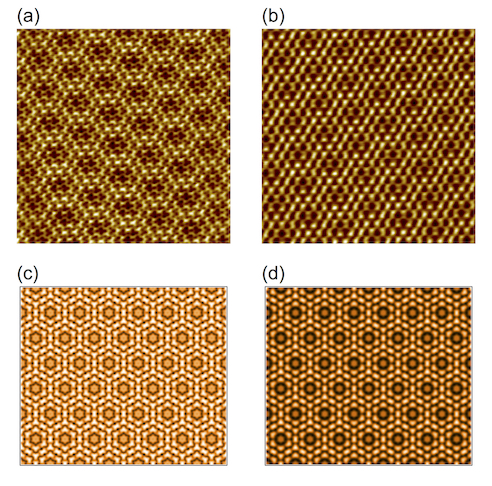}\\
\label{S11_SUPPL_STM_exp_vs_th}
\end{figure}
\noindent\textbf{Fig.\,S11.} Experimental (top row) and calculated (bottom row) STM images of gr/BiAg$_2$-relaxed (system \textbf{C}). Parameters used: (a,c) $U_T=+50$\,mV and (b,d) $U_T=+150$\,mV. Calculations were performed in the framework of the Tersoff-Hamann formalism [J. Tersoff and D. R. Hamann, Phys. Rev. B \textbf{31}, 805 (1985)].  

\end{document}